\documentstyle[twocolumn,epsfig,prb,aps,amsmath,amssymb]{revtex}

\begin{document}

\twocolumn[\hsize\textwidth\columnwidth\hsize\csname
@twocolumnfalse\endcsname

\title{Superconductivity and antiferromagnetism in a hard-core boson
spin-1 model \\ in two dimensions}

\author{Jos\'e A. Riera}
\address{
Instituto de F\'{\i}sica Rosario, Consejo Nacional de
Investigaciones
Cient\'{\i}ficas y T\'ecnicas, y Departamento de F\'{\i}sica,\\
Universidad Nacional de Rosario, Avenida Pellegrini 250,
2000-Rosario, Argentina}
\date{\today}
\maketitle
\begin{abstract}
A model of hard-core bosons and spin-1 sites with single-ion
anisotropy is proposed to approximately describe hole pairs 
moving in a background of singlets and triplets with the aim of
exploring the relationship between superconductivity and 
antiferromagnetism. The properties
of this model at zero temperature were investigated using quantum 
Monte Carlo techniques. The most important feature found 
is the suppression of superconductivity, as long range coherence
of preformed pairs, due to the presence of both antiferromagnetism 
and $S^z=\pm 1$ excitations. Indications of charge ordered and other
phases are also discussed.
\end{abstract}

\smallskip
\noindent PACS: 74.20.-z, 74.25.Dw, 74.25.Ha, 02.70.Uu

\vskip2pc]

The mechanism of pairing and the establishment of long-range 
superconducting (SC) coherence are still central issues in the theory
of high-T$_c$ superconductivity.\cite{kivelson} Although strong 
electronic correlations
are the essential component of most proposed scenarios for the SC phase
and its nearby antiferromagnetic (AF) Mott insulating phase,
important experimental results such as the resonant 
peak\cite{resonant} have not been satisfactorily explained by such 
scenarios. Part of the theoretical limitations are related to the
enormous difficulty in studying a microscopic model like the $t$-$J$
model, which is the simplest model that describes the dynamics of
holes in an AF background. From the numerical point of view, the main
difficulty consists in reaching large enough clusters. This problem
is really critical in the presence of inhomogeneities, like
the well-known stripes\cite{stripes} and the ones that have 
more recently become the center of intensive 
research.\cite{granular}

In this article, we propose and analyze a highly simplified effective
Hamiltonian in two dimensions (2D) to study the interplay between
superconductivity and antiferromagnetism. Our goal is to describe the 
movement of boson hole pairs in a sea of magnetic excitations. This 
model implies a reduction in the Hilbert space with respect to the 
$t$-$J$ model, which is convenient for exact diagonalization 
calculations, and the elimination of the ``minus sign" 
problem\cite{minus-sign} which severely inhibits the application of 
quantum Monte Carlo (QMC) techniques to fermions in 2D. Such a model
could be useful to study experimental features like the ones
mentioned earlier and inhomogeneous states in which there are phase 
separated SC and AF regions. 

Although I shall not attempt a derivation of the present model from
a more microscopic one like the $t$-$J$ model, I shall present some
heuristic arguments to guide the physical interpretation of the
results obtained and shown below. Let us start with the $t$-$J$ model
on the square lattice. A coarse-grained Hamiltonian can be obtained by
mapping two nearest neighbor (NN) sites (a 
``dimer") of the original lattice onto a site of the effective
lattice.\cite{ladder} Due to the constraint of no double-occupancy
in the starting model, there are nine states in each of the coarse-grained 
sites. Since our purpose is to study the interplay between pairs and
magnetic excitations, the states corresponding to only one hole in the
original dimers are eliminated. Hence, we are left with five states per
site which correspond to a singlet ($S=0$), a triplet ($S=1$, and
$S^z=-1, 0,+1$), and a hole pair. Finally, the singlets play the role
of the vacuum, and the hole pairs are described by bosons.\cite{drs}
All these states are by construction hard-core entities.
Of course, this coarse-graining procedure here outlined has been
employed many times, particularly in the context
of resonant valence bond (RVB) theories. By restoring the interactions
between dimers, it has been shown\cite{eder} that the long-range AF
order is recovered in the undoped case.

The physical underlying scenario is then one in which hole pairs
move in a ``soup" of singlets and triplets. The same states per 
site can be also found in ``projected" SO(5) 
models\cite{zhang,dorneich} but in these models
the interactions between them are
essentially dictated by symmetry requirements 
rather than by microscopical considerations 
(see Fig.~\ref{interfig} below). An extreme case with no triplet
excitations, i.e., a model of pairs as hard-core bosons, has
been extensively considered to describe superfluid
phases.\cite{hebert,bernardet,schmid}

However, the model studied in the present work contains a further
simplification and hence it should be considered as 
a ``toy model" of that scenario of pairs moving in a singlet-triplet 
soup. This simplification consists in considering doublets instead 
of triplets. This is achieved by assuming that all sites are occupied
by triplets and pairs, and by adding to the Hamiltonian a single-ion
anisotropy term (third term in Eq.~(\ref{eqham}) below) with a
coupling constant $\Lambda$. For $\Lambda \ne 0$, the triplet states
split into $S^z=\pm 1$ doublets and $S^z=0$ singlets. Of course,
the total spin is no longer a good quantum number. Let us introduce
$D=\sum_i (S_i^z)^2$, the number of doublets. For $\Lambda=0$ there
are only triplets in the model and then $D$ goes to 2/3 of 
the number of triplets. For $\Lambda \rightarrow \infty $ there are
only singlets
and then $D \rightarrow 0$. Assuming that this quantity $D$ evolves
continuously between both limits then one could consider it as
the analog, in this simplified model, of the number of triplet 
excitations in the singlet-triplet soup earlier discussed. In this 
sense, the single-ion anisotropy $\Lambda$ would correspond to the
chemical potential in the related SO(5) models.

The exchange interactions included in our model are those 
resulting from a single exchange in the original spin-1/2
$t$-$J$ model, as shown in Fig.~\ref{interfig}. The exchange
interaction shown in Fig.~\ref{interfig}(b) is not included
in the projected SO(5) Hamiltonian. On the other hand, in that
Hamiltonian, there is a term corresponding to the one shown 
in Fig.~\ref{interfig}(c). However, this interaction has no 
microscopic origin in a $t$-$J$ like model because it does not 
conserve the total $S^z$ and hence it is not
included in our Hamiltonian. Then, the effective Hamiltonian
here proposed is:\cite{note}
\begin{eqnarray}
{\cal H} = &-& t \sum_{\langle i,j\rangle,s} (b_j^\dagger c_{j,s}
c_{i,s}^\dagger b_i + h.c.) + J \sum_{\langle i,j\rangle}
{\bf S}_i \cdot {\bf S}_j \nonumber   \\
&+& \Lambda \sum_i (S_i^z)^2 + V \sum_{\langle i,j\rangle} n_i n_j
\label{eqham}
\end{eqnarray}
\noindent
where $c_{j,s}, c_{j,s}^\dagger$ are annihilation and creation
operators of $s=S^z=0,\pm1$ spins, $b_i, b_i^\dagger$ are annihilation
and creation operators of hole pairs, $n_i=b_i^\dagger b_i$.
The exchange term is just the spin-1 Heisenberg term, and it
captures the corresponding interactions shown in
Fig.~\ref{interfig} in a simple way.

\begin{figure}
\begin{center}
\epsfig{file=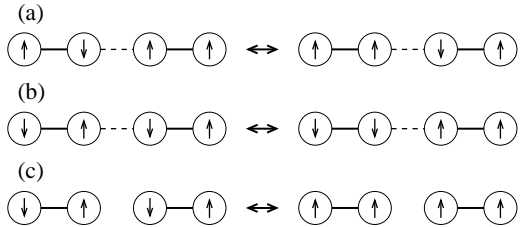,width=6.8cm,angle=0}
\end{center}
\caption{Microscopic origin of exchange interactions. Dimers are
indicated with bold lines. The 
exchange interaction between electrons in different dimers are 
shown with dashed lines.
In terms of the total $S^z$ of each dimer, case (a) corresponds to 
$(0,+1)\leftrightarrow (+1,0)$, and (b) to 
$(0,0)\leftrightarrow (-1,+1)$. The change (c),
$(0,0)\leftrightarrow (+1,+1)$ has no microscopical origin.}
\label{interfig}
\end{figure}

The hopping term between pairs and doublets ($s=\pm 1$ in the first
term of Eq.~(\ref{eqham})) induces the minus sign problem in the
quantum Monte Carlo simulations and hence we shall not include
it in most of the following study. The analogous term of hopping
between pairs and triplets are also not included in the projected
SO(5) Hamiltonian. Preliminary results obtained including this term
show essentially the same qualitative features shown below for the 
parameters considered. Although in principle $\Lambda$ should be
determined by the internal dynamics of the original $t$-$J$ model,
in our effective model this is a parameter which we vary freely,
in the same way as the spin chemical potential in the projected
SO(5) model. Finally, our model contains a nearest neighbor
Coulomb repulsion between pairs, to prevent phase separation in
the low density region.\cite{dorneich}

We adopt $t$ as the unit of energy. From previous studies of the
hard-core boson model we adopt $V=3$. It is not simple to determine
a priori the ratio $J/t$. $J$ should be of the order of the exchange
coupling constant of the original Hamiltonian, although in some
cases it may involve longer than NN interactions.
$t \sim t_0 t_0^\prime/\Delta_b$ where $t_0$ ($t_0^\prime$) is the
NN (longer than NN) hopping amplitude in the original $t$-$J$ model,
and $\Delta_b$ the cost of breaking a hole pair in the intermediate
state. This ratio could be determined eventually a posteriori by
e.g. matching the energies of small clusters of the original and the
effective model. In the present work, we adopted a reasonably small
value $J/t=0.3$. We expect that the effect of a different value would
just amount to quantitative shifting the phase boundaries without
introducing qualitative new features. Of course, a more exhaustive
parameter study should be done in the future.

\begin{figure}
\begin{center}
\epsfig{file=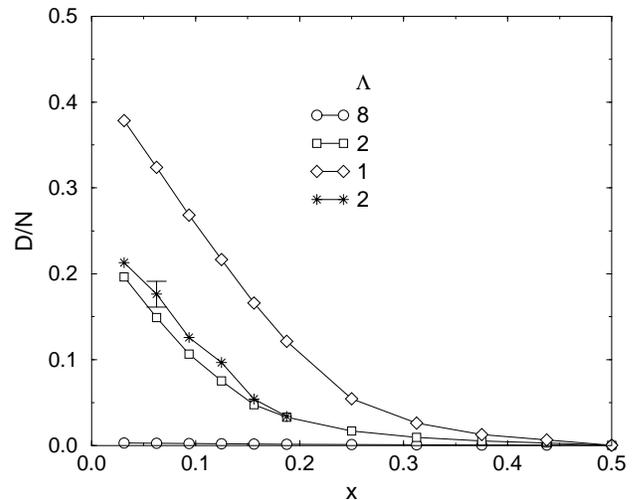,width=6.8cm,angle=-90}
\end{center}
\caption{Relative number of sites with $S^z=\pm 1$ for various values 
of the single-ion anisotropy $\Lambda$. The stars correspond to the
full Hamiltonian Eq.~(\ref{eqham}) at $T=0.2$ (in this case a typical 
error bar is shown).
}
\label{szfig}
\end{figure}

All results shown below, except otherwise stated, correspond to
the model Eq.~(\ref{eqham}) excluding $s=\pm 1$ in the hopping
term, and they were obtained by QMC techniques (conventional
world-line algorithm) on the $8\times 8$
cluster with periodic boundary conditions. The temperature $T$ was
varied between $0.2 t$ and $0.1 t$, i.e., considerably below the
Kosterlitz-Thouless temperatures separating normal and SC phases
found in previous studies on related models.\cite{schmid,dorneich}
The Trotter number was kept at a standard value, $\Delta \tau=0.1$.
The results were finally extrapolated to zero temperature with an
exponential law. The error bars are about or smaller than the
size of the symbols used, except otherwise stated.

The dependence of $D/N$ ($N$: number of sites of the cluster), 
i.e. the relative number of $S^z=\pm 1$ sites
is shown in Fig.~\ref{szfig} as a function of the pair 
density $x$ and for various values of $\Lambda \ge 1$. For $\Lambda=8$, 
$D$ is negligible and hence this value of $\Lambda$ is essentially the
$\Lambda\rightarrow \infty$ limit, i.e., hard-core pairs moving in a
vacuum played by the singlets.\cite{hebert,bernardet,schmid} The 
continuity of $D$ as $\Lambda$ is increased from zero, at a fixed
$x$ is not a trivial problem. It is obvious 
that for $J=0$ there are no terms in the Hamiltonian to compensate 
the cost of the single-ion anisotropy term and hence there is a 
discontinuity in $D$ as soon as $\Lambda$ takes a nonzero value. The 
possibility of a finite critical value of $J$, $J_{cr}(x)$, below
which $D$ is discontinuous with $\Lambda$ will not be examined in the 
present work. It should also be noticed that, after dropping the
$s=\pm 1$ contributions to the first term of Eq.~(\ref{eqham}),
another source of anisotropy appears between the 
$S^z=0, \pm 1$ components. This is reflected in $D$ as it can
be seen in Fig.~\ref{szfig} for $\Lambda=2$.

In order to determine partially the phase diagram of this model
in the $\Lambda-x$ space, we compute the following quantities. In the
first place, charge and spin correlations,
$C({\bf r}) = \langle n_0 n_{\bf r} \rangle$, 
$S({\bf r}) = \langle S^z_0 S^z_{\bf r} \rangle$, respectively, 
together with their Fourier transforms $C({\bf k})$ and 
$S({\bf k})$, i.e., the charge and spin static 
structure factors.
In addition, we compute also the staggered correlations, i.e.,
including a factor $(-1)^{x+y}$ in the sums of the previous 
expressions.

\begin{figure}
\begin{center}
\epsfig{file=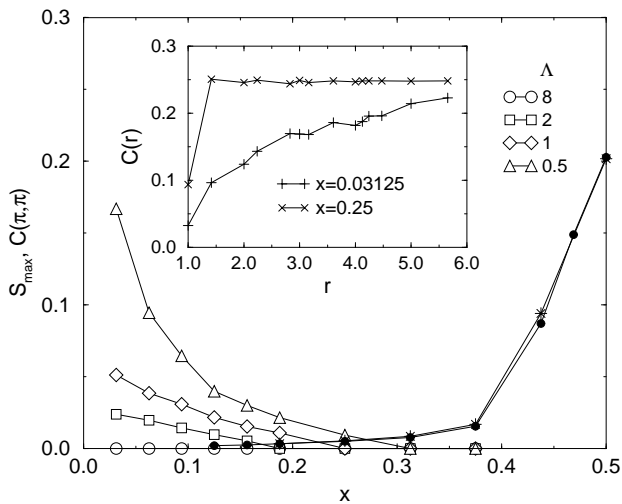,width=6.8cm,angle=-90}
\end{center}
\caption{Staggered spin correlations at the maximum distance (open
symbols) as a function of pair density and for various values of
$\Lambda$. The charge structure factor at $(\pi,\pi )$
is shown for $\Lambda=8$ (circles) and $\Lambda=1$
(stars). In the inset, the charge correlations
vs. distance are shown for $\Lambda=1$ at $x=0.03125$ (multiplied
by 10) and $x=0.25$.
}
\label{affig}
\end{figure}

In Fig.~\ref{affig}, the staggered spin correlations at the maximum
distance $r$, are shown as a function of pair density and for
several values of $\Lambda$. They indicate strong AF correlations 
as $\Lambda\rightarrow 0$, particularly as $x\rightarrow 0$. i.e. as
the pure spin-1 Heisenberg model is approached. The 
determination of the long-range character and of the boundary 
of the AF region as a function of $\Lambda$ would imply a finite 
size scaling, which is out of the scope of the present study.
The charge structure factor has a peak at $(\pi, \pi)$ for all
the parameter space $(x,\Lambda)$ examined except at low densities.
As shown in the Fig.~\ref{affig}, this structure factor
is very weakly dependent with $\Lambda$, and hence the so-called
``checkerboard solid" phase found in the hard-core boson 
limit\cite{hebert} might extend down to $\Lambda=0$ at 
$x\approx 0$ (see discussion below). 

For pair densities lower
than $\approx 0.15$ and for $\Lambda \approx 0$ we have detected 
indications of incommensurate charge ordering with 
$q_{IC}=(\pi-\delta,\pi-\delta)$ below $x=0.06$, and 
$q_{IC}=(\pi-\delta,\pi)$ (and symmetry related points) above it,
with $\delta$ varying with pair density.
In these cases, the charge structure factor is more than two 
orders of magnitude smaller than for $(\pi, \pi)$ at $x=0.5$.

\begin{figure}
\begin{center}
\epsfig{file=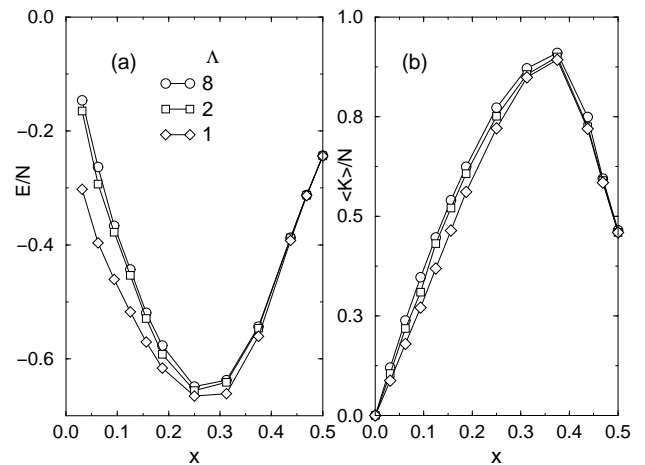,width=6.3cm,angle=-90}
\end{center}
\caption{(a) Energy per site, and (b) kinetic energy per site as 
a function of pair density and for several values of $\Lambda$. 
}
\label{enefig}
\end{figure}

In Fig.~\ref{enefig}(a) the total energy per site vs. pair density 
for various $\Lambda$
is shown. The region $x\ge 0.4$ apparently presents a negative
curvature which corresponds to phase separation (PS), although probably
larger clusters would be needed to confirm this result. In this region,
PS was found for 
$\Lambda\rightarrow \infty$, $0.39\leq x \leq 0.5$ in the bulk 
limit using alternative criteria.\cite{schmid}

Consistently with this indication of PS, pair-pair correlations 
(inset of Fig.~\ref{affig}) show a
tendency of the holes to be at short distances as the density
increases. This tendency is superimposed to a 
strong alternation corresponding to the peak in the charge
structure factor at $(\pi,\pi)$. The second most important
peaks correspond to $q_{IC}=(\pi-\delta,\pi)$, $\delta=\pi/4$ 
(and symmetry related points), and there are no indications 
of pair clustering in a compact region (which would correspond
to $q=(\delta,\delta))$.

\begin{figure}
\begin{center}
\epsfig{file=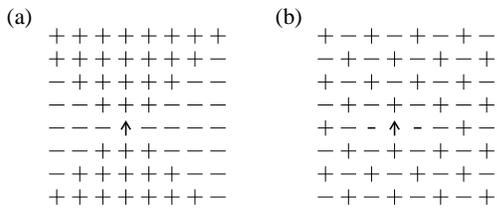,width=6.5cm,angle=0}
\end{center}
\caption{Signs of $C_{{\hat y} {\hat y}}({\bf r})$ ($\hat y$ is the
vertical direction), (a) at the maximum $\chi_{SC}$ and (b) for
$\chi_{SC}\approx 0$, $x\approx 0.5$. The arrows indicate the 
reference bond}
\label{signfig}
\end{figure}

To determine the presence of a superconducting phase, considering
the hard core bosons as tightly-bound Cooper pairs, we 
compute the equal-time current-current correlations:
\begin{eqnarray}
C_{\alpha \alpha}({\bf r}) = \frac{1}{N} \sum_{i} \langle j_{\alpha}(i) 
j_{\alpha}(i+{\bf r}) \rangle
\label{corcur}
\end{eqnarray}
\noindent
where the current operator along 
direction $\alpha$ ($\alpha={\hat x}, {\hat y}$) is:
\begin{eqnarray}
j_{\alpha}(i) = i t \sum_s (b^{\dagger}_{i+\alpha} c_{i+\alpha,s}
c^{\dagger}_{i,s} b_{i}- h.c.)
\label{current}
\end{eqnarray}
\noindent
We adopt as the SC order parameter the correlation
between the total current crossing a border and the current on a 
reference bond at the maximum distance:
\begin{eqnarray}
\chi_{SC} = \sum_x C_{{\hat y} {\hat y}}(x,L/2)
\end{eqnarray}
\noindent
The signs of $C_{{\hat y} {\hat y}}({\bf r})$ ($\hat y$ is the
vertical direction) are shown in Fig.~\ref{signfig}. At the
maximum $\chi_{SC}$ (Fig.~\ref{signfig}(a)), the pattern shows
the presence of current lines extended over all the cluster, while
for $\chi_{SC}\approx 0$, $x\approx 0.5$ (Fig.~\ref{signfig}(b)),
the current loops are of the size of a single plaquette.

\begin{figure}
\begin{center}
\epsfig{file=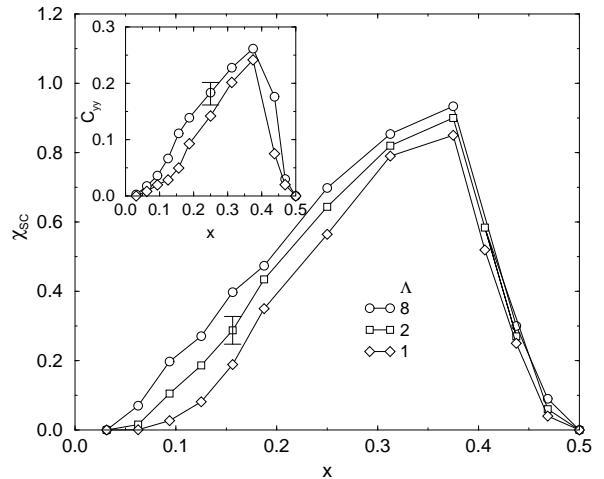,width=6.5cm,angle=-90}
\end{center}
\caption{Superconducting order parameter $\chi_{SC}$ as a function 
of pair density and for several values of $\Lambda$.
The current-current correlations $C_{{\hat y} {\hat y}}$ at the
maximum distance along ${\hat y}$ are shown in the inset
as a function of pair density and for $\Lambda=1$ and 8. Typical
error bars are shown.
}
\label{curfig}
\end{figure}

The most important results obtained with the present model concern
the superconducting order parameter shown in Fig.~\ref{curfig}.
$\chi_{SC}$ and $C_{{\hat y} {\hat y}}({\bf r})$ at the
maximum distance along ${\hat y}$ have a similar behavior.
Even though there are important error bars, there are two apparent
qualitative features. In the first place as $\Lambda$ is reduced
($\Lambda=8$ is representative of the $\Lambda \rightarrow \infty$
behavior) the onset
of the SC phase is pushed to higher values of pair density.
This
feature may be a consequence of AF order at larger densities
although, as said before, a finite size extrapolation should be done
to determine the phase boundaries. Second, the intensity of 
$\chi_{SC}$ is reduced as $\Lambda$ decreases, i.e. as more $S^z=\pm 1$
sites are present in the system. In other words, there is a 
maximal SC signal in the hard-core boson model and this signal is
reduced as the isotropic spin-1 model is approached. 

It is also instructive to compare the behavior of $\chi_{SC}$ 
with the one of the kinetic energy per site shown in 
Fig.~\ref{enefig}(b). Both $\chi_{SC}$ and $\langle K\rangle/N$
decrease with $\Lambda$ for a given $x$, although this effect is more
intense in the former quantity. The suppression of both quantities
near

\noindent
$x=0.5$ may be due to the charge localization in the 
$(\pi,\pi)$ solid phase. Actually, since $\chi_{SC}$ and
$C(\pi,\pi)$ are sizable for $x\ge 0.4$, this suggests a
coexistence of SC and the checkerboard solid in this region,
which corresponds to the PS region found above.\cite{schmid}
On the other hand, the suppression of
SC at low densities has no counterpart in the behavior of the
kinetic energy. The characteristics of this region are hence 
indicating the presence of a AF Mott insulating phase.

\begin{figure}
\begin{center}
\epsfig{file=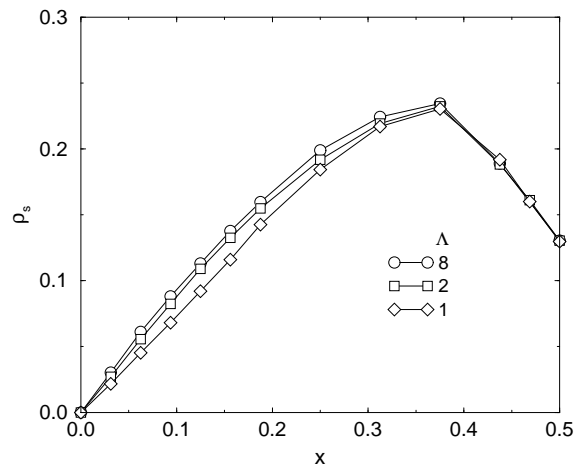,width=6.3cm,angle=-90}
\end{center}
\caption{Superfluid density as a function of pair density and for
several values of $\Lambda$. 
}
\label{supdens}
\end{figure}

Finally in Fig.~\ref{supdens} we show the superfluid density which is
a quantity more conventionally studied in the context of hard-core
bosons. This quantity is computed as\cite{batrouni}
\begin{eqnarray}
\rho_{s} = \int_0^{\beta} \langle j_{\alpha}(\tau) j_{\alpha}
\rangle d\tau
\end{eqnarray}
\noindent
where $\tau$ is the imaginary time, $\beta=1/T$ and 
$j_{\alpha}(\tau) = \sum_{ i}  j_{\alpha}(i,\tau)$.
As it can be seen in this Figure, $\rho_s$ has the same qualitative
behavior as the kinetic energy per site. Hence, as discussed above,
it differs with
$\chi_{SC}$ in the behavior at low pair density.

In summary, a model of pairs as hard-core bosons interacting with
spin-1 sites has been proposed to study the interplay between 
superconductivity and antiferromagnetism. Using quantum Monte Carlo
techniques, its main features as a function of pair density and
single-ion anisotropy have been determined. Even in this ``toy"
model some features, which could have some relevance to high-T$_c$
superconductivity, have been observed. The most important feature
is the suppression of superconductivity, as long range coherence
of preformed pairs, due to the presence of both antiferromagnetism 
and $S^z=\pm 1$ excitations, which in this model represent triplet 
excitations in more realistic models. Indications of incommensurate
charge ordering and phase separation have also been observed.
This model certainly deserves further investigation. In particular,
the dynamical correlations between doublets and pairs should be
examined. Finally,
the proposed model is relatively simple and hence it can be expected
to describe other physical systems. In fact, 
the model can describe a variety of spin-1 systems with
a layered structure like La$_{2-x}$Sr$_x$NiO$_4$ 
(Ref.~\onlinecite{kajimoto}) or with two-dimensionally
coupled chains.\cite{yamamoto} Hole pairs could appear in the former 
compound if there is an on-site attraction present.


\begin{references}

\bibitem{kivelson} V. J. Emery and S. A. Kivelson, Nature {\bf 374},
                 434 (1995).

\bibitem{resonant} P. Bourges, in ``The gap symmetry and Fluctuations
in High Temperature Superconductors", ed. by J. Bok, G. Deutscher, D.
Pavuna and S. A. Wolf (Plenum Press, 1998).

\bibitem{stripes} J. M.Tranquada, J. D. Axe, N. Ichikawa, Y. Nakamura, 
     S. Uchida, and B. Nachumi, Phys. Rev. B {\bf 54}, 7489 (1996);
     N. Ichikawa, S. Uchida, J. M. Tranquada, T. Niemoller, P. M.
     Gehring, S.-H. Lee, and J. R. Schneider, Phys. Rev. Lett. 
     {\bf 85}, 1738 (2000).

\bibitem{granular} K. M. Lang, V. Madhavan, J. E. Hoffman, E. W.
      Hudson, H. Eisaki, S. Uchida, and J. C. Davis,
      cond-mat/001122, to appear in Nature, (2002).

\bibitem{minus-sign} See e.g., S. Chandrasekharan and U.-J. Wiese
     Phys. Rev. Lett. {\bf 83}, 3116 (1999), and references
     therein.

\bibitem{ladder} Similar procedures have been performed on ladders.
     See e.g., R. Eder, A. Dorneich, M. G. Zacher, W. Hanke, and S. C. 
     Zhang, Phys. Rev. B {\bf 59}, 561 (1999); T. Siller, M. Troyer, T.
     M. Rice, and S. R. White, cond-mat/0006080.

\bibitem{drs} It is implicitly assumed that the mechanism of
    pairing is similar to the one found in ladders. See E. Dagotto, J. 
    Riera and D. Scalapino, Phys. Rev. B {\bf 45}, 5744 (1992).

\bibitem{eder} R. Eder, Phys. Rev. B {\bf 59}, 13810 (1999). 

\bibitem{zhang} S. C. Zhang, J-P. Hu, E. Arrigoni, W. Hanke, and
      A. Auerbach, Phys. Rev. B {\bf 60}, 13070 (1999).

\bibitem{dorneich} A. Dorneich, W. Hanke, E. Arrigoni, M. Troyer,
      and S. C. Zhang, cond-mat/0106473.

\bibitem{hebert} F. H\'ebert, G. G. Batrouni, R. T. Scalettar,
       G. Schmid, M. Troyer, and A. Dorneich, Phys. Rev. B {\bf 65},
       014513 (2001).

\bibitem{bernardet} K. Bernardet, G. G. Batrouni, J.-L. Meunier,
      G. Schmid, M. Troyer, and A. Dorneich, cond-mat/0110314.

\bibitem{schmid} G. Schmid, S. Todo, M. Troyer, and A. Dorneich,,
      cond-mat/0110024.

\bibitem{note} A number of models interpolating between the $t$-$J$
    model and the present one, and including details like dimer 
    orientation, can be found in the literature.
    See e.g. Refs.~\onlinecite{ladder,eder}; Ehud Altman and Assa 
    Auerbach, cond-mat/0108087.

\bibitem{batrouni} QMC calculations of the superfluid density have 
      been discussed in e.g., G. G. Batrouni, R. T. Scalettar, and 
      G. T. Zimanyi, Phys. Rev. Lett. {\bf 65}, 1765 (1990).

\bibitem{kajimoto} R. Kajimoto, T. Kakeshita, H. Yoshizawa, T. 
      Tanabe, T. Katsufuji, and Y. Tokura, Phys. Rev. B {\bf 64},
      144432 (2001).

\bibitem{yamamoto} Undoped spin-1 chains with single-ion anisotropy
      have been studied by S. Yamamoto and S. Miyashita, Phys. 
      Rev. B {\bf 50}, 6277 (1994).

\end{references}
\end{document}